\documentclass[a4paper,12pt]{article}
\usepackage{amssymb,amsmath}
\usepackage{epsfig,fancyhdr}

\usepackage{a4}
\usepackage{parskip}
\usepackage{color,soul,rotating}
\usepackage{amsthm,amsxtra,amscd,relsize,multirow}
\usepackage[all,2cell]{xy}\UseAllTwocells

\newcommand{\sect}[1]{\setcounter{equation}{0}\section{#1}}

\def\a{\alpha}
\def\b{\beta}
\def\d{\delta}

\def\s{\sigma}

\def\m{\mu}

\def\f{\phi}
\def\vf{\varphi}

\def\l{\lambda}

\def\o{\omega}
\def\sinh{\mathrm{sinh}}
\def\cosh{\mathrm{cosh}}

\def\p{\partial}
\def\rb{\right}
\def\lb{\left}

\newcommand{\eq}[1]{\begin{equation} #1 \end{equation}}
\newcommand{\al}[1]{\begin{align} #1 \end{align}}
\newcommand{\ml}[1]{\begin{multline} #1 \end{multline}}

\def\cp{\mathbb {CP}^3}

\begin{document}
\begin{titlepage}
\markright{\bf TUW--09--11}
\title{On the pulsating strings in $AdS_4\times \cp$}

\author{H. Dimov${}^{\star}$ and R.~C.~Rashkov${}^{\dagger,\star}$\thanks{e-mail:
rash@hep.itp.tuwien.ac.at.}
\ \\ \ \\
${}^{\star}$  Department of Physics, Sofia
University,\\
5 J. Bourchier Blvd, 1164 Sofia, Bulgaria
\ \\ \ \\
${}^{\dagger}$ Institute for Theoretical Physics, \\ Vienna
University of Technology,\\
Wiedner Hauptstr. 8-10, 1040 Vienna, Austria
}
\date{}
 \end{titlepage}


\maketitle
\thispagestyle{fancy}

 \begin{abstract}
We study the class of pulsating strings on $AdS_4\times{\mathbb CP}^3$. 
Using a generalized ansatz for pulsating string configurations we find
new solutions of this class. Further we quasi-classically quantize the theory
and obtain the first corrections to the energy. The latter, due to AdS/CFT correspondence,
is supposed to give the anomalous dimensions of operators of the gauge theory dual
 ${\cal N}=6$ Chern-Simons theory.
  \end{abstract}

\sect{Introduction}

The attempts to establish a correspondence between the large N
limit of gauge theories and string theory has more than 30 years history and
 over the years it showed different faces. Recently an explicit realizations of 
this correspondence was provided by the Maldacena conjecture about AdS/CFT correspondence
\cite{Maldacena}. The convincing results from the duality between type IIB string
theory on $AdS_5\times S^5$ and ${\cal N}=4$ super Yang-Mills theory
\cite{Maldacena,GKP,Witten}, made this subject  a major research area and many 
fascinating new features have been established.

An important part of the current understanding of the duality between gauge theories and strings (M-theory) 
is the world-volume dynamics of the branes. Recently there has been a number of works focused on the understanding of
the world-volume dynamics of multiple M2-branes - an interest inspired by 
Bagger, Lambert and Gustavsson \cite{Bagger:2006sk} investigations based on the
structure of Lie 3-algebra. 

Recently, motivated by the possible description of the world-volume
dynamics of coincident membranes in M-theory, a new class of
conformal invariant, maximally supersymmetric field theories in 2+1
dimensions has been found \cite{Schwarz:2004yj,ABJM}.
The main feature of these theories is that they contain gauge fields with Chern-Simons-like  kinetic
terms. Based on this development, Aharony, Bergman, Jafferis and
Maldacena proposed a new gauge/string duality between an $\cal N=6$
 super-conformal Chern-Simons theory (ABJM theory) coupled with
bi-fundamental matter, describing $N$ membrane on $S^7/\mathbb Z_k$. This model,
which was first proposed by Aharony, Bergman, Jafferis, and Maldacena
\cite{ABJM}, is believed to be dual to M-theory on $AdS_4\times
S^7/\mathbb Z_k$.

The ABJM theory actually consists of two Chern-Simons theories of level $k$
and $-k$ correspondingly and each with gauge group $SU(N)$. The two pairs of
chiral superfields transform in the bi-fundamental representations
of $SU(N) \times SU(N)$ and the R-symmetry is $SU(4)$ as it should be for
 $\mathcal N=6$ supersymmetry of the theory. It was observed in
\cite{ABJM} that there exists a natural definition of a 't Hooft coupling -- $\lambda = N/k$. 
Ii was observed that in the 't Hooft limit $N\rightarrow \infty$ with $\lambda$ held fixed,
 one has a continuous coupling  $\lambda$ and the ABJM theory is weakly coupled for $\lambda \ll 1$. 
The ABJM theory is conjectured to be dual to M-theory on $AdS_4\times S^7 / \mathbb Z_k$ with
$N$ units of four-form flux. In the scaling limit $N, k \rightarrow\infty$
with $k \ll N \ll k^5$ the theory can be compactified to type IIA string theory on $ads_4 \times \mathbb P^3$.
Thus, the AdS/CFT correspondence, which has led to many exciting developments in the duality between type IIB string
theory on $AdS_5\times S^5$ and ${\cal N}=4$ super Yang-Mills theory, is now being extended to the $AdS_4/CFT_3$
and is expected to constitute a new example of exact gauge/string theory duality.

The semi-classical strings has played, and still play, an important role in studying various aspects of
$AdS_5/SYM_4$ correspondence \cite{Bena:2003wd}-\cite{Lee:2008sk}. The development in this
subject gives a strong hint about how the new emergent duality can be investigated. An important
role in these studies plays the integrability. The superstrings on $AdS_4\times\cp$ as a coset was first
studied in \cite{Arutyunov:2008if}\footnote{See also \cite{Stef}} which opens the door for investigation of the integrable structures in the theory. To pursue these issues it is necessary to have the complete superstring action. It was noticed that
the string supercoset model does not 
describe the entire dynamics of type IIA superstring in AdS4 x CP3, but only its 
subsector. The complete string dual of the ABJM model, i.e. the
complete type-IIA Green-Schwarz string action in  $AdS_4\times\cp$ superspace 
has been constructed in \cite{Gomis:2008jt}.

Various properties on gauge theory side and tests on string theory side  
as rigid rotating strings, pp-wave limit, relation to spin chains, and certain
limiting cases as well as pure spinor formulation  have been considered \cite{Arutyunov:2008if}-\cite{Sundin:2008vt}.

In these intensive studies many properties were uncovered and impressive results obtained,
but still the understanding of this duality is far from completeness.

Another class of string solutions which proved its importance in the case of $AdS_5/CFT_4$ duality
is the class of pulsating strings introduced first in \cite{Minahan:2002rc} and studied and generalized further in
\cite{Engquist:2003rn,Dimov:2004xi,Smedback:1998yn}. In the case of $AdS_4\times\cp$ background the pulsating 
strings are also expected to play an essential role.
The simplest such solutions were mentioned in \cite{Chen:2008qq}, but thorough analysis and quasi-classical quantization
are still missing. The purpose of this paper is to analyze and quasi-classically quantize the class of pulsating strings
on $\cp$ part of the $AdS_4\times\cp$ background. The first corrections to the energy,
which according to the AdS/CFT conjecture gives the anomlaous dimensions of gauge theory operators,  will be the main subject of these considerations.

The paper is organized as follows. In the Introduction we briefly present the basic facts about ABJM theory.
To explain the method and fix the procedure, in the next Section we give short review of the pulsating 
strings and their quasi-classical quantization for the case of $R\times S^3$ background.
Next, we proceed with the pulsating strings on $AdS_4\times\cp$ background restricting the string dynamics to
$\cp$ part of the spacetime. The next section is devoted to the derivation of
the correction to the energy. We find the wave function associated with the Laplace-Beltrami operator
of $\times\cp$ first and then compute the first correction to the energy. The latter is supposed to give the anomalous dimension
of operator of the dual theory. We conclude with a brief discussion on the results.


\paragraph{ABJM and strings on $AdS_4\times \cp$}

To find the ABJM theory one starts with analysis M2-brane dynamics governed by
11d supergravity action \cite{ABJM}
\eq{
S=\frac{1}{2\kappa_{11}^2}\int dx^{11}\s{-g}\left(R-\f{1}{2\cdot
4!}F_{\mu\nu\rho\sigma}F^{\mu\nu\rho\sigma}\right)-\frac{1}{12\kappa_{11}^2}
\int C^{(3)}\wedge F^{(4)}\wedge F^{(4)}, 
\label{abjm-1}
}
 where $\kappa_{11}^2=2^7\pi^8 l_p^9$. Solving for the equations of motions  
\eq{
R^\mu_\nu=\frac{1}{2}\left(\frac{1}{3!}F^{\mu\a\beta\gamma}F_{\nu\a\beta\gamma}
-\frac{1}{3\cdot 4!}\delta^\mu_\nu
F_{\a\beta\rho\sigma}F^{\a\beta\rho\sigma}\right), 
\label{abjm-2}
} 
and 
\eq{
\p_{\sigma}(\sqrt{-g}F^{\sigma\mu\nu\xi})=\frac{1}{2\cdot
(4!)^2}\epsilon^{\mu\nu\xi\a_1\dots\a_8}F_{\a_1\dots\a_4}F_{\a_5\dots\a_8},
\label{abjm-3}
}
one can find the M2-brane solutions whose near horizon limit becomes $AdS_4\times S^7$
\eq{ 
ds^2=\frac{R^2}{4}ds^2_{AdS_4}+R^2 ds^2_{S^7}.
\label{abjm-4}
}
In addition we have $N'$ units four-form flux 
\eq{
 F^{(4)}=\frac{3R^3}{8}\epsilon_{AdS_4}, \quad R=l_p(2^5 N'\pi^2)^{\frac{1}{6}}.
\label{abjm-5}
}
Next we consider the quotient $S^7/\mathbb Z_k$ 
acting as $z_i \rightarrow e^{ i \frac{ 2 \pi}{k } } z_i$. It is convenient
first to write the metric on $S^7$ as
\eq{
ds^2_{S^7} =  ( d \varphi' + \omega)^2 + ds^2_{CP^3},
\label{abjm-6}
}
where
\al{
& ds^2_{CP^3} = \frac{ \sum_i d z_i d \bar z_i}{r^2 } -
    \frac{ | \sum_i z_i d \bar z_i |^2}{r^4} ~,\quad
 r^2 \equiv \sum_{i=1}^4 |z_i|^2,\notag \\
& d \varphi' + \omega  \equiv  \frac{ i}{ 2  r^2 } (  z_i d \bar z_i - \bar z_i d z _i ),\quad
d\omega =  J = { i}   d \left(\frac{ z_i}{ r}\right)  d \left(\frac{ \bar z_i}{ r }\right).
\label{abjm-7}
}
and then to perform the $\mathbb Z_k$ quotient identifying
$\vf'=\vf/k$ with $\vf\sim \vf+2\pi$ ($J$ is proportional to the Kahler form on $\cp$). 
The resulting metric becomes
\eq{
ds^2_{S^7/{\mathbb Z}_k} = \frac{ 1}{ k^2 } ( d \vf+ k \o)^2 + ds^2_{CP^3}.
}
One can observe that the first volume factor on the right hand side is divided by factor of $k$ compared to
the initial one. In order to have consistent quantized flux one must impose $N'=kN$ where $N$ is 
the number of quanta of the flux on the quotient.
We note that the spectrum of the supergravity fields of the final theory is just the projection of the 
initial $AdS_4\times S^7$ onto the $\mathbb Z_k$ invariant states. In this setup there is a natural
definition of {}'t Hooft coupling $\lambda \equiv N/k$. Decoupling limit should be taken
as $N,k\rightarrow \infty$ while $N/k$ is kept fixed.

One can follow now \cite{ABJM} to make reduction to type IIA with the following final result
\al{
 ds^2_{string} = & \frac{ R^3}{ k} ( \frac{ 1 }{ 4 } ds^2_{AdS_4} + ds^2_{{\mathbb CP}^3 } ),
 \\
 e^{2 \phi} = & \frac{ R^3}{k^3 } \sim \frac{ N^{1/2}}{ k^{5/2} }= \frac{ 1 }{ N^2 } 
\left( \frac{ N}{ k } \right)^{5/2},
 \\
 F_{4} = & \frac{ 3}{ 8 }  {  R^3}  \epsilon_4 ,
 \quad
 F_2 =  k d \omega = k J,
 }
 We end up then with  $AdS_4 \times\cp$ compactification
 of type IIA string theory with $N$ units of $F_4$ flux on $AdS_4$ and $k$ units of 
$F_2$ flux on the ${\mathbb CP}^1 \subset\cp$ 2-cycle.

The radius of curvature in string units is $R^2_{str} = \frac{ R^3}{ k}  =  2^{5/2} \pi 
\sqrt{ \lambda}$.
It is important to note that the type IIA approximation is valid in the regime where $k \ll N \ll k^5$.


In order to fix the notations, we write down the explicit form of the metric on $AdS_4\times\mathbb{CP}^3$
in spherical coordinates. The metric on $AdS_4\times\mathbb{CP}^3$ can be written as \cite{PopeWarner}
\begin{multline}
 ds^2=R^2\left\lbrace \dfrac{1}{4}\left[ -\cosh^2\rho\,dt^2+d\rho^2+\sinh^2\rho\,d\Omega_2^2\right] \right.\\
\left.+d\mu^2+\sin^2\mu\left[d\alpha^2+\dfrac{1}{4}\sin^2\alpha(\sigma_1^2
+\sigma_2^2+\cos^2\alpha\sigma_3^2)+\dfrac{1}{4}\cos^2\mu(d\chi+\sin^2\mu\sigma_3)^2\right]\right\rbrace.
\label{metric}
\end{multline}
Here $R$ is the radius of the $AdS_4$, and $\sigma_{1,2,3}$ are left-invariant 1-forms on an $S^3$, parameterized by $(\theta,\phi,\psi)$,
\begin{align}
&\sigma_1=\cos\psi\,d\theta+\sin\psi\sin\theta\,d\phi,\notag\\
&\sigma_2=\sin\psi\,d\theta-\cos\psi\sin\theta\,d\phi,\label{S3}\\
&\sigma_3=d\psi+\cos\theta\,d\phi.\notag
\end{align}
The range of the coordinates is
$$0\leq\mu,\,\alpha\leq\dfrac{\pi}{2},\,\,0\leq\theta\leq\pi,\,\,0\leq\phi\leq2\pi,\,\,0\leq\chi,\,\psi\leq4\pi.$$


\sect{A brief review of pulsating strings on $R\times S^3$}

In this section we give a brief review of the pulsating string
solutions obtained first by Minahan \cite{Minahan:2002rc} and generalized later
in \cite{Engquist:2003rn,Dimov:2004xi,Smedback:1998yn}. We will concentrate only on the case of pulsating
string on $S^5$ part of $AdS_5\times S^5$, i.e.  a circular
string which pulsates expanding and contracting on $S^5$ is considered. To fix the
notations, we start with the metric of $S^5$ and relevant part of
$AdS_5$
\eq{
ds^2=R^2\lb(\cos^2\theta d\Omega_3^2+d\theta^2+\sin^2\theta d\psi^2+
d\rho^2-\cosh^2 dt^2\rb),
\label{1.1}
}
where $R^2=2\pi{\a}^\prime\sqrt{\lambda}$. In order to obtain the simplest
solution, we identify the target space time with the worldsheet
one, $t=\tau$, and use the ansatz $\psi=m\sigma$, i.e. the string is
stretched along $\psi$ direction, $\theta=\theta(\tau)$ and keep the
dependence on $\rho=\rho(\tau)$ for a while. The reduced Nambu-Goto
action in this case is
\eq{
S=m\sqrt{\lambda}\int dt \sin\theta \sqrt{\cosh^2\rho-\dot\theta^2}.
\label{1.1a}
}
For our considerations it is useful to pass to Hamiltonian
formulation. For this purpose we find the canonical momenta and find
the Hamiltonian in the form\footnote{For more details see \cite{Minahan:2002rc}.}
\eq{
H=\cosh\rho\sqrt{\Pi_\rho^2+\Pi_\theta^2+m^2\lambda\sin^2\theta}.
\label{1.2}
}
Fixing the string to be at the origin of $AdS_5$ space ($\rho=0$), we
see that the squared Hamiltonian have a form very similar to a point
particle. The last term in (\ref{1.2}) can be considered as a
perturbation, so first we find the wave function for a free particle
in the above geometry
\ml{
\frac{\cosh\rho}{\sinh^3\rho}\dfrac{d}{d\rho}\cosh\rho \sinh^3\rho
\dfrac{d}{d\rho}\Psi(\rho,\theta)-\frac{\cosh^2\rho}{\sin\theta
  \cos^3\theta} \dfrac{d}{d\theta}\sin\theta\,\cos^3\theta
\dfrac{d}{d\theta} \Psi(\rho,\theta)\\
=E^2\Psi(\rho,\theta).
\label{1.3}
}
The solution to the above equation is
\eq{
\Psi_{2n}(\rho,\theta)=(\cosh\rho)^{-2n-4}\, P_{2n}(\cos\theta)
\label{1.4}
}
where $P_{2n}(\cos\theta)$ are spherical harmonics on $S^5$ and the
energy is given by
\eq{
E_{2n}=\Delta=2n+4.
\label{1.5}
}
Since we consider highly excited states (large energies), one should
take large $n$, so we can approximate the spherical harmonics as
\eq{
P_{2n}(\cos\theta)\approx \sqrt{\frac{4}{\pi}}\cos(2n\theta).
\label{1.6}
}
The correction to the energy can be obtained by using perturbation
theory, which to first order is
\eq{
\delta E^2=\int\limits_0^{\pi/2}d\theta\,
\Psi_{2n}^\star(0,\theta)\,m^2\lambda \sin^2\theta\,
\Psi_{2n}(\theta)
=\frac{m^2\lambda}{2}.
\label{1.7}
}
Up to the first order in $\lambda$ we find for the anomalous dimension of
the corresponding YM operators\footnote{See \cite{Minahan:2002rc} for more
  details.}
\eq{
\Delta-4=2n[1+\frac 12\,\frac{m^2\lambda}{(2n)^2}].
\label{1.8}
}
It should be noted that in this case the $R$-charge is zero. In order to
include it, we consider pulsating string on $S^5$ which has a center
of mass that is moving on the $S^3$ subspace of $S^5$ \cite{Engquist:2003rn,Dimov:2004xi,Smedback:1998yn}. 
While in the previous example $S^3$ part of the metric was assumed trivial, now we
consider all the $S^3$ angles to depend on $\tau$ (only). The
corresponding Nambu-Goto action now is
\eq{
S=-m\sqrt{\lambda}\int\,
dt\sin\theta\,\sqrt{1-\dot\theta^2-\cos^2\theta
  g_{ij}\dot\phi^i\dot\phi^j },
\label{1.9}
}
where $\phi_i$ are $S^3$ angles and $g_{ij}$ is the corresponding
$S^3$ metric. The Hamiltonian in this case is \cite{Engquist:2003rn,Dimov:2004xi,Smedback:1998yn}
\eq{
H=\sqrt{\Pi_\theta^2+\frac{g^{ij}\Pi_i\Pi_j}{\cos^2\theta}
  +m^2\lambda\sin^2\theta}.
\label{1.10}
}
Again, we see that the squared Hamiltonian looks like the point
particle one, however, now the potential has angular
dependence. Denoting the quantum number of $S^3$ and $S^5$ by $J$ and
$L$ correspondingly, one can write the Schrodinger equation 
\eq{
-\frac{4}{\omega}\dfrac{d}{d\omega}\Psi(\omega)
+\frac{J(J+1)}{\omega}\Psi(\omega) = L(L+4) \Psi(\omega),
\label{1.11}
}
where we set $\omega=\cos^2\theta$. The solution to the Schrodinger
equation is
\eq{
\Psi(\omega)=\frac{\sqrt{2(l+1)}}{(l-j)!}\,\frac{1}{\omega}\left(
  \frac{d}{d\omega}\right)^{l-j}
\omega^{l+j}(1-\omega)^{l-j}
, \quad j=\frac{J}{2}; l=\frac{L}{2}.
\label{1.12}
}
The first order correction to the energy $\delta E$ is
\eq{
\delta E^2=m^2\lambda\,\frac{2(l+1)^2-(j+1)^2-j^2}{(2l+1)(2l+3)},
\label{1.13}
}
or, up to first order in $\lambda$
\eq{
E^2=L(L+4)+m^2\lambda\frac{L^2-J^2}{2L^2}
\label{1.14}
}
The anomalous dimension then is given by
\eq{
\gamma=\frac{m^2\lambda}{4L}\a(2-\a),
\label{1.15}
}
where $\a=1-J/L$.

We conclude this section referring for more details to \cite{Minahan:2002rc} and
\cite{Engquist:2003rn,Dimov:2004xi,Smedback:1998yn}.


\sect{Pulsating strings on $AdS_4\times\cp$}
In this section we consider a circular pulsating string expanding and contracting only on $\mathbb{CP}^3$ part of $AdS_4\times\mathbb{CP}^3$. We will consider the string dynamics confined in the $\cp$ part of the spacetime with
 no extension of the string in the remaining spatial coordinates. Then the relevant metric 
we will work with, is given by
\begin{equation}
ds^2_{t\times\mathbb{CP}^3}=R^2\left(-\frac{1}{4}dt^2+ ds^2_{\mathbb{CP}^3}\right),\quad
R^2=\sqrt{32\pi^2\lambda}. 
\end{equation}
Having in mind the explicit form of the $\mathbb{CP}^3$ metric \eqref{metric}, one  can 
write it as
\begin{equation}
ds^2_{\mathbb{CP}^3}=g_{ij}dx^i dx^j+\hat{g}_{pq}dy^p dy^q,
\end{equation}
where  the part $g_{ij}$ is defined as
\begin{equation}
g_{ij}=diag\left( 1,\,\sin^2\mu,\,\frac{1}{4}\sin^2\mu\sin^2\alpha\right),\quad i,j=1,\,2,\,3,\quad
x^1=\mu,\,\,x^2=\alpha,\,\,x^3=\theta,
\end{equation} 
while $\hat{g}_{pq}=\hat{g}_{pq}(\mu,\,\alpha,\,\theta)$ is the remaining part of metric associated with 
$\phi,\,\psi,\,\chi$ coordinates, denoted here as $p,\,q=1,\,2,\,3,\quad y^1=\phi,\,\,y^2=\psi,\,y^3=\chi$.
The residual worldsheet symmetry allows us to identify $t$ with $\tau$ and to obtain a classical pulsating string solution. To do that we use the following ansatz 
\begin{align}
&x^1=x^1(\tau)=\mu(\tau),&x^2=x^2(\tau)=\alpha(\tau),&\quad x^3=x^3(\tau)=\theta(\tau),\\
&y^1=\phi=m_1\sigma+h^1(\tau),&y^2=\psi=m_2\sigma+h^2(\tau),&\quad y^3=\chi=m_3\sigma+h^3(\tau).
\end{align}
We are interested in the induced worldsheet metric which in our case has the form
\begin{equation}
ds^2_{ws}=R^2\left\lbrace\left(-\frac{1}{4}+ g_{ij}\dot{x}^i \dot{x}^j+\hat{g}_{pq}
\dot{h}^p \dot{h}^q\right)d\tau^2 +(\hat{g}_{pq}m_p m_q)d\sigma^2+
2(\hat{g}_{pq}m_p \dot{h}^q)d\tau d\sigma \right\rbrace.
\end{equation} 
The Nambu-Goto action
\begin{equation}
 S_{NG}=-T\,\int\,d\tau\,d\sigma
\sqrt{-det(G_{\mu\nu}\partial_\alpha\,X^{\mu}\partial_\beta\,X^{\nu})},
\end{equation} 
in this ansatz then reduces to expression 
\begin{equation}
 S_{NG}=-TR^2\int d\tau d\sigma
\sqrt{\left(\frac{1}{4}- g_{ij}\dot{x}^i \dot{x}^j-\hat{g}_{pq}\dot{h}^p \dot{h}^q\right)(\hat{g}_{pq}m_p m_q)+
(\hat{g}_{pq}m_p \dot{h}^q)(\hat{g}_{pq}m_q \dot{h}^p)},
\end{equation} 
where $TR^2=2\sqrt{2\lambda}$, and we set for simplicity $\quad\alpha'=1$.
To follow the procedure described in the previous Section, we need a Hamiltonian formulation of the problem.
For this purpose, we have to find first the canonical momenta of our dynamical system.
Straightforward calculations give for the momenta
\begin{equation}
\Pi_i=2\sqrt{2\lambda}\dfrac{(\hat{g}_{pq}m_p m_q)g_{ij} \dot{x}^j}{\sqrt{\left(\frac{1}{4}- g_{ij}\dot{x}^i \dot{x}^j-\hat{g}_{pq}\dot{h}^p \dot{h}^q\right)(\hat{g}_{pq}m_p m_q)+
(\hat{g}_{pq}m_p \dot{h}^q)(\hat{g}_{pq}m_q \dot{h}^p)}},\quad i=1,2,3,
\end{equation} 
\begin{equation}
\hat{\Pi}_p=2\sqrt{2\lambda}\dfrac{(\hat{g}_{pq}m_p m_q)\hat{g}_{pq}\dot{h}^q-
(\hat{g}_{pq}m_p \dot{h}^q)\hat{g}_{pq}m_q}
{\sqrt{\left(\frac{1}{4}- g_{ij}\dot{x}^i \dot{x}^j-\hat{g}_{pq}\dot{h}^p 
\dot{h}^q\right)(\hat{g}_{pq}m_p m_q)+
(\hat{g}_{pq}m_p \dot{h}^q)(\hat{g}_{pq}m_q \dot{h}^p)}},\quad p=1,2,3.
\end{equation} 
Solving for the derivatives in terms of the canonical momenta and substituting back into 
the Legendre transform of the Lagrangian, we find the Hamiltonian 
\begin{equation}
H^2=\frac{1}{4}\left( g^{ij}\Pi_i \Pi_j+\hat{g}^{pq}\hat{\Pi}_p\hat{\Pi}_q\right)+
\frac{(2\sqrt{2\lambda})^2(\hat{g}_{pq}m_p m_q)}{4}.
\end{equation} 
The interpretation of this relation is as in the case of $AdS_5\times S^5$, namely, the
first term in the brackets is the kinetic term while the second one is considered as 
a potential $V$, which in our case has the form
\begin{equation}
V(\mu,\,\alpha,\,\theta)=2\lambda\,\hat{g}_{pq}(\mu,\,\alpha,\,\theta)m_p m_q.
\end{equation} 
The approximation where our considerations are valid assumes high energies, 
which suggests that one can think of this potential term as a perturbation.
For latter use we write down the explicit form of the potential 
\begin{multline}
V(\mu,\,\alpha,\,\theta)=\frac{\lambda}{2} \left\lbrace
\sin^2\mu \sin^2\alpha\left[ \sin^2\theta+\cos^2\alpha\cos^2\theta(1+\sin^2\alpha)\right]\,m_1^2+\right.\\
+\sin^2\mu \sin^2\alpha\cos^2\alpha(1+\sin^2\alpha)\,m_2^2+\sin^2\mu\cos^2\alpha \,m_3^2+\\
+2\sin^2\mu \sin^2\alpha\cos^2\alpha\cos\theta(1+\sin^2\alpha)\,m_1 m_2+\\
+2\sin^2\mu \sin^2\alpha\cos^2\alpha\cos\theta \,m_1 m_3+\\
\left.+2\sin^2\mu \sin^2\alpha\cos^2\alpha \,m_2 m_3
\right\rbrace .
\label{potential}
\end{multline}
The above perturbation to the free action will produce the corrections to the energy 
and therefore to the anomalous dimension. In order to calculate the corrections to the energy 
as a perturbations due to the above potential however,
we need the normalized wave function associated with $\cp$ space and all 
these issues are subject to the next section.


\sect{Semiclassical correction to the energy}

In this Section we will compute the quasi-classical correction to the energy of the
pulsating string on $\cp$. As we discussed in the previous Section, the Hamiltonian
of the pulsating string is interpreted as a dynamical system with high energy
described by free theory Schr\"odinger equation and perturbed by a potential
\eqref{potential}.

First of all, we have to find the wave function associated with the  
Laplace-Beltrami operator on $\cp$ and then to obtain the correction to the energy 
due to the induced potential.

\subsection{The $\cp$ metric and the associated Laplace-Beltrami operator}

We start with the Laplace-Beltrami operator in coordinates parameterizing
 $\mathbb{CP}^3$ as follows. First we define coordinates
\begin{align}
& y_1+i\,y_2=\mu\sin\alpha\cos\frac{\theta}{2}\,e^{\frac{i}{2}(\phi+\psi+\chi)}\notag\\
& y_3+i\,y_4=\mu\sin\alpha\sin\frac{\theta}{2}\,e^{\frac{i}{2}(\phi-\psi-\chi)}
\label{parametr}\\
& y_5+i\,y_6=\mu\cos\alpha\,e^{\frac{i}{2}\chi}\notag.
\end{align}
The line element of $\cp$ in these coordinates is explicitly given by:
\begin{align}
ds^2_{\mathbb{CP}^3}=& d\mu^2+\sin^2\mu d\alpha^2+\dfrac{1}{4}\sin^2\mu \sin^2\alpha d\theta^2
+\dfrac{1}{4}\sin^2\mu \sin^2\alpha\left[ \sin^2\theta  \right.\notag \\
& \left. +\cos^2\alpha\cos^2\theta(1+\sin^2\alpha)\right]d\phi^2+
+\dfrac{1}{4}\sin^2\mu \sin^2\alpha\cos^2\alpha(1+\sin^2\alpha)d\psi^2 \notag \\
& +\dfrac{1}{4}\sin^2\mu\cos^2\alpha d\chi^2
+\dfrac{2}{4}\sin^2\mu \sin^2\alpha\cos^2\alpha\cos\theta(1+\sin^2\alpha)d\psi d\phi  \notag\\
& +\dfrac{2}{4}\sin^2\mu \sin^2\alpha\cos^2\alpha\cos\theta d\chi d\phi
+\dfrac{2}{4}\sin^2\mu \sin^2\alpha\cos^2\alpha d\psi d\chi 
\label{metric-a}
\end{align}

\paragraph{The Laplas-Beltrami operator on $\mathbb{CP}^3$} \,\,
Using the standard definition of the Laplace-Beltrami operator in global coordinates we find
\begin{multline}
\bigtriangleup(\mu,\,\alpha,\,\theta,\,\phi,\,\psi,\,\chi)=\dfrac{1}{\sin^5\mu}
\,\dfrac{\partial}{\partial\mu}\left[\sin^5\mu\,\dfrac{\partial}{\partial\mu} \right]
+\dfrac{1}{\sin^2\mu\sin^3\alpha\cos^2\alpha}\,\dfrac{\partial}{\partial\alpha}
\left[ \sin^3\alpha\cos^2\alpha\,\dfrac{\partial}{\partial\alpha}\right] \\
+\dfrac{4}{\sin^2\mu\sin^2\alpha\sin\theta}\,\dfrac{\partial}{\partial\theta}\left[ \sin\theta\dfrac{\partial}{\partial\theta}\right] 
+\dfrac{4}{\sin^2\mu\sin^2\alpha\sin^2\theta}\,\dfrac{\partial^2}{\partial\phi^2}
-\dfrac{8\cos\theta}{\sin^2\mu\sin^2\alpha\sin^2\theta}\,
\dfrac{\partial^2}{\partial\phi\partial\psi}  \\
+\dfrac{4(\sin^2\theta+\cos^2\alpha\cos^2\theta)}{\sin^2\mu\sin^2\alpha\sin^2\theta\cos^2\alpha}\,
\dfrac{\partial^2}{\partial\psi^2}-\dfrac{8}{\sin^2\mu\cos^2\alpha}\,
\dfrac{\partial^2}{\partial\chi\partial\psi}  
+\dfrac{4(1+\sin^2\alpha)}{\sin^2\mu\cos^2\alpha}\,\dfrac{\partial^2}{\partial\chi^2}
\end{multline}
Since we are going to separate the variables, it is useful to arrange the terms as follows
\begin{multline}
\bigtriangleup(\mu,\,\alpha,\,\theta,\,\phi,\,\psi,\,\chi)=\dfrac{1}{\sin^5\mu}
\,\dfrac{\partial}{\partial\mu}\left(\sin^5\mu\,\dfrac{\partial}{\partial\mu} \right) \\
+\dfrac{1}{\sin^2\mu}\,\left\lbrace\,\dfrac{1}{\sin^3\alpha\cos^2\alpha}\,
\dfrac{\partial}{\partial\alpha}\left( \sin^3\alpha\cos^2\alpha\,\dfrac{\partial}{\partial\alpha}\right)\right.\\
+\dfrac{4}{\sin^2\alpha\cos^2\alpha}\,\left(\dfrac{\partial}{\partial\psi}
-\sin^2\alpha\dfrac{\partial}{\partial\chi}\right)^2
+\dfrac{4}{\cos^2\alpha}\,\dfrac{\partial^2}{\partial\chi^2}\\
\left. +\dfrac{4}{\sin^2\alpha}
\left[\dfrac{1}{\sin\theta}\,\dfrac{\partial}{\partial\theta}\left( 
 \sin\theta\dfrac{\partial}{\partial\theta}\right)
+\dfrac{1}{\sin^2\theta}\left( \dfrac{\partial}{\partial\phi}
-\cos\theta\,\dfrac{\partial}{\partial\psi}\right)^2\right]\right\rbrace.
\end{multline}
Then the operator $\bigtriangleup(\mu,\,\alpha,\,\theta,\,\phi,\,\psi,\,\chi)=\dfrac{1}{\sin^5\mu}$ can be written as:
\begin{equation}
\bigtriangleup(\mu,\,\alpha,\,\theta,\,\phi,\,\psi,\,\chi)=\dfrac{1}{\sin^5\mu}
\,\dfrac{\partial}{\partial\mu}\left(\sin^5\mu\,\dfrac{\partial}{\partial\mu} \right)
+\dfrac{1}{\sin^2\mu}\,\bigtriangleup(\alpha,\,\theta,\,\phi,\,\psi,\,\chi),
\end{equation}
where
\begin{multline}
\bigtriangleup(\alpha,\,\theta,\,\phi,\,\psi,\,\chi)=
\dfrac{1}{\sin^3\alpha\cos^2\alpha}\,\dfrac{\partial}{\partial\alpha}\left( \sin^3\alpha\cos^2\alpha\,\dfrac{\partial}{\partial\alpha}\right)+\\
+\dfrac{4}{\sin^2\alpha\cos^2\alpha}\,\left(\dfrac{\partial}{\partial\psi}
-\sin^2\alpha\dfrac{\partial}{\partial\chi}\right)^2
+\dfrac{4}{\cos^2\alpha}\,\dfrac{\partial^2}{\partial\chi^2}
+\dfrac{4}{\sin^2\alpha}\,\bigtriangleup(\theta,\,\phi,\,\psi)
\end{multline}
and
\begin{equation}
\bigtriangleup(\theta,\,\phi,\,\psi)=\dfrac{1}{\sin\theta}\,
\dfrac{\partial}{\partial\theta}\left(\sin\theta\dfrac{\partial}{\partial\theta}\right)
+\dfrac{1}{\sin^2\theta}\left( \dfrac{\partial}{\partial\phi}
-\cos\theta\,\dfrac{\partial}{\partial\psi}\right)^2.\label{LB}
\end{equation}
The full measure on $\mathbb{CP}^3$ is
\begin{equation}
d\Omega(\mu,\,\alpha,\,\theta)=\frac{1}{16}\sin^5\mu\sin^3\alpha\cos^2\alpha\sin\theta 
d\mu d\alpha d\theta.\label{measure}
\end{equation} 
\paragraph{The wave function}\,\,

The Schr\"{o}dinger equation for the wave function is
\begin{equation}
\bigtriangleup(\mu,\,\alpha,\,\theta,\,\phi,\,\psi,\,\chi)\,\Psi(\mu,\,\alpha,
\,\theta,\,\phi,\,\psi,\,\chi)=-4E^2\,\Psi(\mu,\,\alpha,\,\theta,\,\phi,\,\psi,\,\chi).
\end{equation}
To separate the variables, we define $\Psi$ as
$\Psi(\mu,\,\alpha,\,\theta,\,\phi,\,\psi,\,\chi)=f(\mu)f(\alpha)f(\theta)f(\phi,\,\psi,\,\chi)$,
where 
$$f(\phi,\,\psi,\,\chi)=\exp(il_1\phi)\exp(il_2\psi)\exp(il_3\chi),\,\,\,\,l_1,\,l_2,\,l_3\in\mathbb{Z}.
$$ 
With this choice
we can solve for the eigenfunctions replacing the derivatives along Killing directions
($\p_\phi,\,\p_\psi,\,\p_\chi$) by ($il_1,\,il_2,\,il_3$) correspondingly.
The equation for $f(\theta)$ then separates from the rest and has the form
\begin{multline}
\left[\bigtriangleup(\theta,\,\phi,\,\psi)
+L^2\right]\,f(\theta) \\ \equiv
\left\lbrace \dfrac{1}{\sin\theta}\,
\dfrac{d}{d\theta}\left(\sin\theta\dfrac{d}{d\theta}\right)
-\dfrac{1}{\sin^2\theta}\left(l_1-\cos\theta\,l_2\right)^2+L^2\right\rbrace\,f(\theta)=0.
\end{multline}
It is convenient to define a new variable $z=\cos\theta$ in which the equation can be written as
\begin{equation}
\left\lbrace
(1-z^2)\,\dfrac{d^2}{dz^2}-2z\,\dfrac{d}{dz}
-\dfrac{1}{1-z^2}\left(l_1-z\,l_2\right)^2+L^2\right\rbrace\,f(z)=0.
\end{equation}
The solution to this equation is
\begin{multline}
f(z)=(1-z)^{\frac{\mid l_1-l_2\mid}{2}}\,(1+z)^{\frac{\mid l_1+l_2\mid}{2}}\,\times\\
{ }_2F_1\left[\dfrac{1}{2}\left(\mid l_1-l_2\mid+\mid l_1+l_2\mid +1 -\sqrt{1+4(l_2^2+L^2)}\right)\,,\,\right.\\
\left.\dfrac{1}{2}\left(\mid l_1-l_2\mid+\mid l_1+l_2\mid +1+\sqrt{1+4(l_2^2+L^2)} \right)\,;\,
1+\mid l_1+l_2\mid\,,\,
\dfrac{1+z}{2}\right]
\end{multline}
Our wave function must be square integrable with respect to the measure on $\cp$ \eqref{measure}.
In separated variables this means that $f(\theta)$ must be square integrable with respect to the measure
on $\theta$. This condition imposes the following restriction on the parameters
\begin{equation}
\sqrt{1+4(l_2^2+L^2)}-\mid l_1-l_2\mid-\mid l_1+l_2\mid -1 =\,2n,\,\,\,\,n\in\mathbb{N}
\end{equation}
Introducing new parameters $\alpha=\mid l_1+l_2\mid$ and $\beta=\mid l_1-l_2\mid$ the solution can be written in the form
\begin{equation}
f(z)=(1-z)^{\alpha/2}\,(1+z)^{\beta/2}\,\dfrac{n!\,\Gamma(\alpha+1)}{\Gamma(\alpha+1+n)}
P^{(\alpha,\beta)}_n(z).\label{sol-z}
\end{equation}
The equation for $f(\alpha)$ can be treated analogously and has the form
\begin{multline}
\lb[\bigtriangleup(\alpha,\,\theta,\,\phi,\,\psi,\,\chi)
+M^2\rb]\,f(\alpha)\\
\equiv 
\left\lbrace \dfrac{1}{\sin^3\alpha\cos^2\alpha}\,\dfrac{d}{d\alpha}\left( \sin^3\alpha\cos^2\alpha\,
\dfrac{d}{d\alpha}\right)
-\dfrac{4}{\sin^2\alpha\cos^2\alpha}\,\left(l_2-\sin^2\alpha\,l_3\right)^2\right.\\
\left. -\dfrac{4}{\cos^2\alpha}\,l_3^2
-\dfrac{4}{\sin^2\alpha}\,L^2+M^2\right\rbrace \,f(\alpha)=0.
\end{multline}
Withe the help of the new variable $y=\sin^2\alpha$, the above equation can be written as
\begin{equation}
\left\lbrace
4y(1-y)\,\dfrac{d^2}{dy^2}+2(4-7y)\,\dfrac{d}{dy}
-\dfrac{4}{y(1-y)}\left(l_2-y\,l_3\right)^2-\dfrac{4}{1-y}\,l_3^2
-\dfrac{4}{y}\,L^2+M^2\right\rbrace\,f(y)=0.
\end{equation}
Recognizing the hypergeometric equation above we find the solution as
\begin{multline}
f(y)=y^{\frac{\sqrt{1+4(L^2+l_2^2 )}-1}{2}} (1-y)^{2((l_2-l_3)^2+l_3^2)} \times \\
{}_2F_1 \left[\dfrac{1}{2}\left( \frac{3}{2}+\sqrt{1+4(L^2+l_2^2)}+4((l_2-l_3)^2+l_3^2)-
\dfrac{\sqrt{25+16(M^2+4l_3^2)} }{2}\right)\,,\right.\\
\,\,\,\,\left.\dfrac{1}{2}
\left( \frac{3}{2}+\sqrt{1+4(L^2+l_2^2)}+4((l_2-l_3)^2+l_3^2)+
\dfrac{\sqrt{25+16(M^2+4l_3^2)}}{2}\right)\,;\right.\\
\left.\, 1+\sqrt{1+4(L^2+l_2^2)} \,,\,y \right]
\end{multline}

In addition we have to ensure that the solution $f(\alpha)$ is square integrable with respect to the measure 
in $\alpha$. This requirement imposes the following restriction on the parameters
\begin{equation}
\dfrac{\sqrt{25+16(M^2+4l_3^2)}}{2}
-\frac{3}{2}-\sqrt{1+4(L^2+l_2^2)}-4((l_2-l_3)^2+l_3^2)=\,2m,\,\,\,\,m\in\mathbb{N}.
\end{equation}
It is useful to introduce new parameters $p=\sqrt{1+4(L^2+l_2^2)}$ and $q=\frac{1}{2}+4((l_2-l_3)^2+l_3^2)$,
 in terms of which the solution can be written in the form
\begin{equation}
f(y)=y^{(p-1)/2} (1-y)^{(q-\frac{1}{2})/2}
\,\dfrac{m!\,\Gamma(p+1)}{\Gamma(p+1+m)}
P^{(p,q)}_m(1-2y).\label{sol-y}
\end{equation}
Then, the equation for $f(\mu)$ we have to solve becomes
\begin{equation}
\left\lbrace \dfrac{1}{\sin^5\mu}\dfrac{d}{d\mu}\left(\sin^5\mu \dfrac{d}{d\mu}\right)-\dfrac{M^2}{\sin^2\mu}
+4E^2\right\rbrace f(\mu)=0.
\end{equation} 
We change the variable defining  $x=\cos\mu$
\begin{equation}
\left\lbrace
(1-x^2)\,\dfrac{d^2}{dx^2}-6x\,\dfrac{d}{dx}
-\dfrac{1}{1-x^2} M^2+4E^2\right\rbrace\,f(x)=0.
\end{equation}
The solution is
\begin{equation}
f(x)=(1-x^2)^{-1}\,LegendreP\left[ \dfrac{\sqrt{25+16E^2}-1}{2}\,;\,\sqrt{4+M^2}\,,\,x\right]
\end{equation}
It is convenient to define new parameters $k=\dfrac{\sqrt{25+16E^2}-1}{2},\quad k=2,3,\dots$ 
and $l=\sqrt{4+M^2},\quad l=2,3,\dots$ and write the solution in the form
\begin{equation}
f(x)=(1-x^2)^{-1}\,P_k^l(x).\label{sol-x}.
\end{equation}
In terms of the new variables the measure \eqref{measure} becomes
\begin{equation}
d\Omega(x,\,y,\,z)=\frac{1}{16}(1-x^2)^2\,\frac{y(1-y)^{1/2}}{2}dxdydz,\quad
0\leq x,\,y\leq 1,\quad -1 \leq z\leq 1.
\end{equation} 
The normalized wave functions of \eqref{sol-z}, \eqref{sol-y}, \eqref{sol-x} have a form
\begin{equation}
\Psi_n^{\alpha,\beta}(z)=\sqrt{\frac{(2n+\alpha+\beta+1)n!\Gamma(n+\alpha+\beta+1)}
{2^{\alpha+\beta+1}\,\Gamma(n+\alpha+1)\Gamma(n+\beta+1)}}
(1-z)^{\alpha/2}\,(1+z)^{\beta/2}P^{(\alpha,\beta)}_n(z),\label{wfun-z}
\end{equation} 

\begin{equation}
\Psi_m^{p,q}(y)=\sqrt{\frac{2(2m+p+q+1)m!\Gamma(m+p+q+1)}{\Gamma(m+p+1)\Gamma(m+q+1)}}
y^{(p-1)/2} (1-y)^{(q-\frac{1}{2})/2}
\,P^{(p,q)}_m(1-2y),\label{wfun-y}
\end{equation}

\begin{equation}
\Psi_k^l (x)=4\,\sqrt{\frac{(2k+1)(k-l)!}{(k+l)!}}\,(1-x^2)^{-1}\,P_k^l(x).\label{wfun-x}
\end{equation}

\subsection{Corrections to the energy}

Potential \eqref{potential} in terms $x,\,y,\,z$:
\begin{multline}
V(x,\,y,\,z)=\frac{\lambda}{2} (1-x^2)\left\lbrace m_1^2 y(1-z^2)+y(1-y^2)\left[
m_1^2 z^2+2m_1 m_2 z+m_2^2\right]+\right.\\
\left.+m_3^2 (1-y)+2y(1-y)\left[ m_1 m_3 z+m_2 m_3\right]\right\rbrace.
\end{multline}
Correction to the energy:
\begin{equation}
\delta E^2=\int\limits_0^1 \int\limits_0^1 \int\limits_{-1}^1 d\Omega(x,\,y,\,z)\,
V(x,\,y,\,z) \Psi_k^{l \,^2}(x) \Psi_m^{p,q\,^2}(y) \Psi_n^{\alpha,\beta\,^2}(z).
\end{equation}

The explicit form of the above formula using the explicit form of the various wave functions is
\begin{multline}
\delta E^2=\frac{\lambda}{2} \dfrac{1}{16}\int\limits_0^1 (1-x^2)\Psi_k^{l \,^2}(x)(1-x^2)^2 dx\times\\
\times\left\lbrace \,m_1^2\,\frac{1}{2}\int\limits_0^1 y\Psi_m^{p,q\,^2}(y) y(1-y)^{1/2} dy
\int\limits_{-1}^1 (1-z^2)\Psi_n^{\alpha,\beta\,^2}(z) dz+\right.\\
\left.+\frac{1}{2}\int\limits_0^1 y(1-y^2)\Psi_m^{p,q\,^2}(y) y(1-y)^{1/2} dy\times\right.\\
\left.\times\left[m_1^2\,\int\limits_{-1}^1 z^2\Psi_n^{\alpha,\beta\,^2}(z) dz+
2m_1 m_2\, \int\limits_{-1}^1 z\Psi_n^{\alpha,\beta\,^2}(z) dz+m_2^2\right] +\right.\\
\left.+m_3^2\,\frac{1}{2}\int\limits_0^1 (1-y)\Psi_m^{p,q\,^2}(y) y(1-y)^{1/2} dy+
\frac{1}{2}\int\limits_0^1 y(1-y)\Psi_m^{p,q\,^2}(y) y(1-y)^{1/2} dy\times\right.\\
\left.\times\left[m_1 m_3\, \int\limits_{-1}^1 z\Psi_n^{\alpha,\beta\,^2}(z) dz+m_2 m_3\right] \right\rbrace ,
\end{multline}

In short notations it looks as
\eq{
\d E^2=\frac{\l}{2}I_0\big\{
m_1^2 I_1.I_2+I_5.\big[m_1^2(1-I_2)+2m_1m_2 I_6+m_2^2\big]
+m_3^2 I_3+ I_4\big[m_1m_3 I_6+m_2m_3\big]\big\},
\label{corr-ener-cp3}
}
where the integrals $I_k$ are explicitly calculated in the Appendix. The expression for the correction to the energy looks very complicated. Therefore, to make conclusions we use the fact that the approximation we work in is for large quantum numbers, say $n,m\gg (p,q), (\a,\b) \gg 1,2$ and
$k\gg l \gg 1,2$. Within this approximation the integrals behave like
\begin{equation}
I_0=\frac{1}{2}+\frac{1}{2}\left(l^2-\frac{1}{4}\right)\,\dfrac{1}{k^2}+O\left(\frac{1}{k^3}\right)
\end{equation}
\eq{
I_1=\frac{1}{2}+\frac{1}{8}\left(p^2-q^2\right)\,\dfrac{1}{m^2}+O\left(\frac{1}{m^3}\right)
}
\eq{
I_2=\frac{1}{8}+\frac{1}{32}\left(2\alpha^2+2\beta^2-1\right)\,
\dfrac{1}{n^2}+O\left(\frac{1}{n^3}\right)
}
\eq{
I_3=\frac{1}{2}+\frac{1}{8}\left(q^2-p^2\right)\,\dfrac{1}{m^2}+O\left(\frac{1}{m^3}\right)
}
\eq{
I_4=\frac{1}{8}+\frac{1}{32}\left(2 p^2+2 q^2-1\right)\,\dfrac{1}{m^2}+O\left(\frac{1}{m^3}\right)
}
\eq{
I_5=\frac{3}{36}+\frac{1}{64}\left(5 p^2+7 q^2-3 \right)\,\dfrac{1}{m^2}+O\left(\frac{1}{m^3}\right)
}
\eq{
I_6=1+\frac{1}{4}\left(\beta^2-\alpha^2\right)\,\dfrac{1}{n^2}+O\left(\frac{1}{n^3}\right)
}

Ignoring the terms of high order ($\frac{p^2}{m^2}\frac{\alpha^2}{n^2},\,\frac{q^2}{m^2}\frac{\alpha^2}{n^2},\, \frac{p^2}{m^2}\frac{\beta^2}{n^2},\,\frac{q^2}{m^2}\frac{\beta^2}{n^2},\,
\frac{1}{m^2}\frac{1}{n^2}....$) we obtain:
\begin{multline}
\delta E^2 \approx \frac{\lambda}{2}\left[\frac{1}{2}+\frac{1}{2}
\left(l^2-\frac{1}{4}\right)\,\dfrac{1}{k^2}\right]\times\\
\times \left\lbrace \frac{29}{128}m_1^2+\frac{3}{16}m_2^2+\frac{1}{2}m_3^2+\frac{3}{8}m_1 m_2+\frac{1}{8}m_1 m_3+\frac{1}{8}m_2 m_3\right.\\
-\left[\frac{21}{512}m_1^2+\frac{3}{64}m_2^2+\frac{3}{32}m_1 m_2
+\frac{1}{32}m_1 m_3+\frac{1}{32}m_2 m_3\right]\,\dfrac{1}{m^2}
-\frac{5}{512}m_1^2\,\dfrac{1}{n^2}\\
+\left[\frac{43}{512}m_1^2+\frac{5}{64}m_2^2-\frac{1}{8}m_3^2
+\frac{5}{32}m_1 m_2+\frac{1}{16}m_1 m_3+\frac{1}{16}m_2 m_3\right]\,
\dfrac{p^2}{m^2}\\
+\left[\frac{41}{512}m_1^2+\frac{7}{64}m_2^2+\frac{1}{8}m_3^2
+\frac{7}{32}m_1 m_2+\frac{1}{16}m_1 m_3+\frac{1}{16}m_2 m_3\right]\,\dfrac{q^2}{m^2}\\
+\left[\frac{5}{256}m_1^2-\frac{3}{32}m_1 m_2-\frac{1}{32}m_1 m_3\right]\,\dfrac{\alpha^2}{n^2}\\
\left. +\left[\frac{5}{256}m_1^2+\frac{3}{32}m_1 m_2+\frac{1}{32}m_1 m_3\right]\,\dfrac{\beta^2}{n^2}\right\rbrace .
\end{multline}

We see that the corrections to the energy have complicated form even at this level. Nevertheless, one can recognize some structures of the quantum numbers that appear in the energy corrections. In order to make comparison with the much well known case of pulsating strings, we present below the case of pulsating strings on ${\mathbb CP}^1$.


\paragraph{Pulsating in ${\mathbb CP}^1$} \

The case of pulsating on ${\mathbb CP}^1$ subspace of the spacetime can be defined as follows.
The metric can be written as
\eq{
ds^2_{\cp}=d\m^2+\sin^2\m\lb[\frac{1}{4}\cos^2\m(\d\chi+\sin^2\m\,\sigma_3)^2+ds^2_{{\mathbb CP}^2}\rb]
}
where
\eq{
ds^2_{{\mathbb CP}^2}=d\a^2+\frac{1}{4}\sin^2\a\Big[
\underset{ds^2_{{\mathbb CP}^1}}{\underbrace{\sigma_1^2+\sigma_2^2}}+\cos^2\a\,\sigma_3^2\Big],
\label{cp2-metric}
}
and as before $\sigma_{1,2,3}$ are left-invariant 1-forms on an $S^3$.
Since $\a=|l_1+l_2|$ and $\b=|l_1-l_2|$ it follows that $\a=\b=|l_1|$ and
$$
L^2=(n+|l_1|)\left[(n+|l_1|)+1 \right]=(n+\a)\left[(n+\a)+1 \right]\sim (n+\a)^2=(n+|l_1|)^2.
$$
The potential in this case becomes
\eq{
V=\frac{\l}{2}m_1^2\sin^2\theta=\frac{\l}{2}m_1^2 (1-z^2).
}
The correction to the energy can be computed using the same technology as before and it gives
the expression:
\begin{equation}
\delta E^2_{{\mathbb CP}^1}=\frac{\l}{2}m_1^2\int\limits_{-1}^1 (1-z^2)\Psi_n^{\alpha,\alpha\,^2}(z) dz.
\label{corr-ener-cp1}
\end{equation}
Ti obtain an explicit expression we have to calculate the following integral
\begin{multline}
I_2^{\a\b}=\int\limits_{-1}^1 (1-z^2)\Psi_n^{\alpha,\beta\,^2}(z) dz=\\
=\frac{(n+\alpha+\beta+1)(n+\alpha+\beta+2)(n+\alpha+1)(n+\beta+1)}{(2n+\alpha+\beta+1)(2n+\alpha+\beta+2)^2(2n+\alpha+\beta+3)}+\\
+\dfrac{n(n+\alpha+\beta+1)}{(2n+\alpha+\beta+1)^2}\left( \frac{n+\alpha+1}{2n+\alpha+\beta+2}-\frac{n+\beta}{2n+\alpha+\beta}\right)^2+\\
+\frac{n(n-1)(n+\alpha)(n+\beta)}{(2n+\alpha+\beta+1)(2n+\alpha+\beta)^2(2n+\alpha+\beta-1)}.
\end{multline}
In \eqref{corr-ener-cp1}  $\a=\b=|l_1|$ and therefore we edn up with
\begin{equation}
I_2^{\a\a}=\frac{1}{16}\frac{(n+2\alpha+1)(n+2\alpha+2)}{(n+\alpha+1)(n+\alpha+3)}
+\frac{1}{16}\frac{n(n-1)}{(n+\alpha+1)(n+\alpha-1)}.
\end{equation}
On other hand we need the limit of large $L$  
$$
L \sim (n+\a)=(n+|l_1|),\,\,\,\, n=L-\a,
$$
which gives the following approximation of the integral $I_2^{\a\a}$
\begin{multline}
I_2=\frac{1}{16}\frac{(L+\alpha+1)(L+\alpha+2)}{(L+1)(L+3)}
+\frac{1}{16}\frac{n(n-1)}{(L+1)(L-1)}\sim\\
\sim \frac{1}{16}\left(1+\frac{\a}{L} \right)^2+\frac{1}{16}\frac{(L-\a)^2-L+\a}{L^2-1}\sim
\dfrac{1}{8}\left[1+\left( \frac{\a}{L}\right)^2+\frac{\a}{2L^2} \right] 
\end{multline}
Substituting into the expression for the correction to the energy we find
\begin{equation}
\delta E^2_{{\mathbb CP}^1}=\frac{\l}{2}m_1^2\dfrac{1}{8}\left[1+\left( \frac{|l_1|}{L}\right)^2+\frac{|l_1|}{2L^2} \right].
\end{equation}
We note that in the limit of large $L$ we reproduce the of \cite{Engquist:2003rn}. 


\sect{Conclusions}

In this section we summarize the results of our study. The purpose of
this paper was to look for pulsating string solutions in $AdS_4\times\cp$
background. The class of pulsating strings has been used to study AdS/CFT correspondence 
in the case of $AdS_5\times S^5$ \cite{Minahan:2002rc,Dimov:2004xi,Smedback:1998yn} and the corrections to the string energy has been associated with anomalous dimensions of certain type operators.

The recently suggested duality between string theory in $AdS_4\times\cp$
attracted much attention promising to be exact and with number of applications.
Here we consider a generalized string ansatz for pulsating string in the $\cp$ part of the geometry.
Next we
considered the corrections to the classical energy. From AdS/CFT point of
view the corrections to the classical energy give the anomalous dimensions of
the operators in SYM theory and therefore they are of primary interest.

For this purpose, we consider the Nambu-Goto actions and find
the Hamiltonian. After that we quantize the resulting theory semi-classically
and obtain the corrections to the energy. Since we consider highly excited
system, the kinetic term is dominating. All this means that we effectively
perform summation over all classical solutions (not only those we explicitly
found) while the effective potential term serves for a small perturbation.
The obtained corrections to the classical energy look complicated, but in certain limit one can find relatively simple expressions. To identify the contribution of the different terms, it is instructive to look at the solutions for the $\mathbb{CP}^1$ case. 
At the end of the last Section we present these solutions thus providing a basis for comparison.
Since they correspond to a subsector of the well known from $AdS_5\times S^5$ considerations
one can identify the origin of the various contribution. 
One can see that the corrections to the energy have 
analogous structure to the case of pulsating, say in $S^5$. The mixing between quantum numbers of the different isometry directions shows up in analogous, but slightly more complicated way. This can be seen using the result from $\mathbb{CP}^1$ subsector and its embedding in $\cp$. 

As a final comment, we note that in order to complete the analysis from AdS/CFT
point of view, it is of great interest to perform an analysis 
comparing our result to that in SYM side. We leave this important question
for future research.

\section*{Acknowledgments}
This work was supported in part by the Austrian Research Fund FWF I192,
NSFB VU-F-201/06 and DO 02-257.

\begin{appendix}

\sect{Some integrals appearing the the main text}

The integrals appeared in \eqref{corr-ener-cp3} are defined as follows:
\ml{
I_0=\dfrac{1}{16}\int\limits_0^1 (1-x^2)\Psi_k^{l \,^2}(x)(1-x^2)^2 dx \\
=1-\frac{(k-l+1)(k+l+1)}{(2k+1)(2k+3)}-\frac{(k-l)(k+l)}{(2k+1)(2k-1)},
}
\begin{multline}
I_1=\frac{1}{2}\int\limits_0^1 y\Psi_m^{p,q\,^2}(y) y(1-y)^{1/2} dy \\
=\frac{(m+p)(m+p+q)}{(2m+p+q+1)(2m+p+q)}+\frac{(m+1)(m+q+1)}{(2m+p+q+1)(2m+p+q+2)},
\end{multline}

\begin{multline}
I_3=\frac{1}{2}\int\limits_0^1 (1-y)\Psi_m^{p,q\,^2}(y) y(1-y)^{1/2} dy \\
=\frac{(m+q)(m+p+q)}{(2m+p+q+1)(2m+p+q)}+\frac{(m+1)(m+p+1)}{(2m+p+q+1)(2m+p+q+2)},
\end{multline}

\begin{multline}
I_4=\frac{1}{2}\int\limits_0^1 y(1-y)\Psi_m^{p,q\,^2}(y) y(1-y)^{1/2} dy \\
=\frac{(m+p+q+1)(m+p+q+2)(m+p+1)(m+q+1)}{(2m+p+q+1)(2m+p+q+2)^2(2m+p+q+3)}+\\
+\dfrac{m(m+p+q+1)}{(2m+p+q+1)^2}\left( \frac{m+p+1}{2m+p+q+2}-\frac{m+q}{2m+p+q}\right)^2+\\
+\frac{m(m-1)(m+p)(m+q)}{(2m+p+q+1)(2m+p+q)^2(2m+p+q-1)},
\end{multline}

\begin{multline}
I_5=\frac{1}{2}\int\limits_0^1 y(1-y^2)\Psi_m^{p,q\,^2}(y) y(1-y)^{1/2} dy \\
=I_1-\left[\frac{(m+p+q+1)(m+p+q+2)(m+p+q+3)(m+p+3)(m+p+2)(m+p+1)}{(2m+p+q+1)(2m+p+q+2)^2
(2m+p+q+3)^2(2m+p+q+4)}+\right.\\
\left.+9\,\frac{(m+p+q+1)(m+p+q+2)(m+q)m(m+p+2)(m+p+1)}{(2m+p+q+1)(2m+p+q+2)
(2m+p+q+3)^2(2m+p+q)^2}+\right.\\
\left.+9\,\frac{(m+p+q+1)(m+q-1)(m+q)m(m-1)(m+p+1)}{(2m+p+q+1)(2m+p+q+2)^2
(2m+p+q-1)^2(2m+p+q)}+\right.\\
\left.+\frac{(m+q-2)(m+q-1)(m+q)m(m-1)(m-2)}{(2m+p+q+1)(2m+p+q)^2
(2m+p+q-1)^2(2m+p+q-2)}\right],
\end{multline}

\begin{multline}
I_2=\int\limits_{-1}^1 (1-z^2)\Psi_n^{\alpha,\beta\,^2}(z) dz \\
=\frac{(n+\alpha+\beta+1)(n+\alpha+\beta+2)(n+\alpha+1)(n+\beta+1)}{(2n+\alpha+\beta+1)
(2n+\alpha+\beta+2)^2(2n+\alpha+\beta+3)}+\\
+\dfrac{n(n+\alpha+\beta+1)}{(2n+\alpha+\beta+1)^2}\left( \frac{n+\alpha+1}{2n+\alpha+\beta+2}-\frac{n+\beta}{2n+\alpha+\beta}\right)^2+\\
+\frac{n(n-1)(n+\alpha)(n+\beta)}{(2n+\alpha+\beta+1)(2n+\alpha+\beta)^2(2n+\alpha+\beta-1)},
\end{multline}

\begin{equation}
\int\limits_{-1}^1 z^2\Psi_n^{\alpha,\beta\,^2}(z) dz=1-I_2,
\end{equation} 

\begin{multline}
I_6=\int\limits_{-1}^1 z\Psi_n^{\alpha,\beta\,^2}(z) dz \\
=\frac{2(n+\beta)(n+\alpha+\beta)}{(2n+\alpha+\beta+1)(2n+\alpha+\beta)}+
\frac{2(n+1)(n+\alpha+1)}{(2n+\alpha+\beta+1)(2n+\alpha+\beta+2)}.
\end{multline}

\end{appendix}


\end{document}